\documentclass[twocolumn,preprintnumbers,amsmath,amssymb,showkeys]{revtex4}
\usepackage{epsfig,latexsym}
\usepackage{graphicx}
\usepackage{dcolumn}
\usepackage{bm}
\usepackage{longtable}
\def\Caption[#1][#2][#3]{\caption{\label{#1}\small{{\bf #2.} #3}}}
\def\ddt[#1]{\frac{d#1(t)}{dt}}
\def\partialddt[#1]{\frac{\partial#1}{\partial t}}
\def\formula[#1][#2]{\begin{equation}\label{#1}#2\end{equation}}
\def\eq[#1]{equation~\ref{#1}}
\def\Eq[#1]{Equation~\ref{#1}}
\def\eqPair[#1][#2]{equations~\ref{#1} and \ref{#2}}
\def\EqPair[#1][#2]{Equations~\ref{#1} and \ref{#2}}
\def\eqs[#1][#2]{equation~\ref{#1} to \ref{#2}}
\def\Eqs[#1][#2]{Equation~\ref{#1} to \ref{#2}}
\def\cfeq[#1]{(cf. equation~\ref{#1})}
\def\tab[#1]{table~\ref{#1}}
\def\Tab[#1]{Table~\ref{#1}}
\def\cftab[#1]{(cf. table~\ref{#1})}
\def\sec[#1]{section~\ref{#1}}
\def\Sec[#1]{Section~\ref{#1}}
\def\cfsec[#1]{(cf. section~\ref{#1})}
\def\fig[#1]{figure~\ref{#1}}
\def\Fig[#1]{Figure~\ref{#1}}
\def\cffig[#1]{(cf. figure~\ref{#1})}
\def\bs{$\!\!$}
\def\img[#1]{{\scriptsize {\bf #1}}}
\def\size[#1]{(size #1$\times$#1 pixels)}

\def\pics{./}
\begin{document}
\textheight=9in
\textwidth = 6.5in	
\voffset=0.75in
\def\imgwidth{0.5}
%
\providecommand\new[1]{{#1}}
%
%
%
%
\title[Adaptation With Dynamically Switching Gain Control]{Pushing it to the Limit:
Adaptation With Dynamically Switching Gain Control}
%
\author{Matthias S. Keil}
\thanks{This manuscript (co-authored with Jordi Vitri{\`a}) has been submitted to \textit{EURASIP Journal on Applied Signal Processing} - Special Issue on Image Perception, to appear 3rd Quarter 2006}
\email{threequarks@yahoo.com}
\affiliation{%
Centre de Visi{\'o} per Computador,
Edifici O,
Campus UAB,
E-08193 Bellaterra (Cerdanyola),
Barcelona (Spain)}%
\affiliation{Computer Science Department and
Centre de Visi{\'o} per Computador}%
%
%
%
\date{\today}
\begin{abstract}
With this paper we propose a model to simulate the functional aspects of light
adaptation in retinal photoreceptors.  Our model, however, does not link specific
stages to the detailed molecular processes which are thought to mediate adaptation
in real photoreceptors.  We rather model the photoreceptor as a self-adjusting
integration device, which adds up properly amplified luminance signals.  The
integration process and the amplification obey a switching behavior that acts to locally
shut down the integration process in dependence on the internal state of the receptor.
The mathematical structure of our model is quite simple, and its computational
complexity is quite low.  We present results of computer simulations which
demonstrate that our model adapts properly to at least four orders of input magnitude.   
\end{abstract}
%
%
%
\maketitle
\normalsize
%
\section{Introduction}
%
There is agreement that adaptation (i.e. the adjustment of sensitivity) is
important for the function of nervous systems, since without corresponding
mechanisms any neuron with its limited dynamic range would stay silent
or operate in saturation most of the time \cite{WalravenEtAl90}. Because neurons
are noisy devices, reliable information transmission is only granted if the
distribution of levels in the stimulus matches the neuron's reliable operation
range \cite{BarlowLevick76}.\\
Consider, for example, the mammalian visual system, with the retina at its front-end.
When performing saccades, the retina must cope with intensity variations which may
span about one \cite{HoodFinkelstein86,ManteEtAl05} to about two orders of magnitude
(2 including shadows acoording to \cite{HoodFinkelstein86}, 2-3 according to \cite{vanHateren1997}).
From one scene to another  (e.g., from bright sunlight to starlight), the range of intensity
variations may well span up to ten orders of magnitude \cite{Martin83,ShapEnCug84,Laughlin89,NormannEtAl91}).
This range of intensities has to be mapped onto less than two orders of output
activity of retinal ganglion cells \cite{Barlow81}, implying some form of
compression of the scale of intensity values.  The retina achieves this by
making use of a cascade of gain control and adaptation mechanisms, respectively
(e.g., \cite{HoodReview98,MeisterBerryII99,FahrenfortEtAl99,FainEtAlReview01}).
Specifically, cone photoreceptors may decrease their sensitivity proportionally
to background intensity, over about 8 log units of background intensity \cite{Burkhardt94}.
This relationship is known as Weber's law (e.g., \cite{Dowling1987}).  Adaptation
in photoreceptors \footnote{Many of the data were gained from rod photoreceptors because
they are more amenable to analysis.  It is generally believed, however, that similar processes
are also taking place in cones.} is achieved by subtly balanced network of molecular processes
(see \cite{KolbFernandezNelson00} for an excellent introduction, and
\cite{BurnsBaylor01,FainEtAlReview01} with references).\\
With the present paper, we propose a mechanism which mimics the dark and light
adaptation of retinal cones.  Our mechanism abstracts from the detailed molecular
processes of the transduction cascade \new{as described in the following section}. 
\new{We seeked out} an easy
implementable and computationally efficient way of achieving the adaptation
behavior of cone photoreceptors.  Our approach should -- and will be -- contrasted with
the retinal stage of a recently proposed model of Grossberg and Hong \cite{GrossHong03,GrossHong04},
which simulates \textit{(i)} luminance adaptation at the outer segment of
the photoreceptor (c.f. \cite{CarpenterGrossberg}), and \textit{(ii)} inhibition
at the inner segment of the photoreceptor by horizontal cells
(e.g.,  \cite{KamermansEtAl01}).  In their model, horizontal cells are
coupled with gap junctions (forming a syncytium), whose connectivity or
permeability decreases with increasing differences between the input of
adjacent cells \cite{Lamb76,PiccolinoEtAl84}.  In other words, their
horizontal cell network establishes current flows inside of regions that
are defined by low contrasts, whereas no activity exchange occurs between
regions which are separated by high contrast boundaries (very similar to
an anisotropic diffusion mechanism \cite{PeronaMalik90}).  In this way,
contrast adaption is implemented.  Notice that our model lacks the latter
stage, and only simulates the photoreceptor adaptation.\\
%
%
\section{Mechanisms of adaptation in the retina}\label{adaptation}
%
A response to light is initiated by photoisomerization of the chromophore 11-\textit{cis}-retinal
to all-\textit{trans}-retinal.  In darkness, 11-\textit{cis}-retinal is bound to rhodopsin in
its inactive conformation, and lies buried in the membranes of the outer segment discs.
Upon absorption of a photon, and the subsequent photoisomerization of the chromophore,
the rhodopsin undergoes a conformational change which converts it into its active form
Rh$^{*}$ (or metarhodopsin II).  The presence of Rh$^{*}$ triggers two distinct mechanisms:
an recycling process known as visual cycle, and an enzymatic cascade known as transduction
cascade.\\
The visual cycle begins with the phosphorylation of Rh$^{*}$, and subsequent binding of
arrestin to the phosphorylated photopigment.  After binding of arrestin, the photopigment
is rendered completely inactive.  The protein opsin is then dephosphorylated, and
all-\textit{trans}-retinal is reduced to all-\textit{trans}-retinol.  The retinol
is isomerized to the 11-\textit{cis}-isomer outside the photoreceptor (in the adjacent
retinal pigment epithelium layer), and re-enters afterwards to re-combine with the
dephosphorylated opsin.\\
The transduction cascade begins with the serial activation of transducins by Rh$^{*}$,
implementing the first stage for signal amplification in the cascade \cite{CalvertEtAl02}.
Thereby, an active complex T$_{\alpha}\cdot$GTP is formed, which binds to and activates
the enzyme phosphodiesterase (PDE).  PDE reduces the concentration of cytoplasmatic cGMP
by hydrolizing it.  The latter process constitutes a second stage for amplification.
The hydrolysis of cGMP causes the closing of cGMP-gated channels, what in turn
generates the electrical response of photoreceptors.  Thus, photoreceptors are
depolarized in darkness because of their open cationic channels, and get hyperpolarized
by light.
\def\sodium{Na$^{+}$}
\def\calcium{Ca$^{2+}$ }
\def\exchanger{Na$^{+}$/K$^{+}$-\calcium\bs-exchanger}
In darkness, the steady current that flows into the outer segment is usually called
dark- or circulating current \footnote{The photocurrent is brought back to the dark-adapted
level by hydrolizing the GTP to GDP.}.
The main fraction of the circulating current is carried by \sodium ions, and a smaller
fraction of \calcium ions \cite{PerryMcNaughton91}.  Calcium is transported out of the
outer segment by the Na$^{+}$/K$^{+}$-\calcium-exchange protein at a constant rate,
independent of the light hitting the photoreceptor.  This implies that light decreases
intracellular \calcium levels, because of the increased probability of channel
opening.  As a consequence, a direct correlation (i.e., a linear relationship)
exists between the circulating current and \calcium concentration. \\ 
Adaptation of the photoreceptor to ambient light is granted by balancing the just
described amplification mechanisms (for low light situations) against mechanisms
which prevent response saturation (e.g., for sunlit scenes).  This balance is
implemented by feedback mechanisms which act either on the catalytic activity
or on the catalytic lifetime of the components that make up the phototransduction
cascade \cite{CalvertEtAl98}.  It is now well established that changes in \calcium
 concentration regulate the cascade in at least three important ways.\\
First, \calcium can prolong the lifetime of Rh$^{*}$ through the inhibition of phosphorylation
in the visual cycle, by means of recoverin.  Second, in the transduction cascade, \calcium 
regulates the  cytoplasmatic concentration of cGMP by binding to guanylate cyclase -- the enzyme
that is responsible for cGMP synthesis.  Third, decreasing \calcium concentrations increase
the sensitivity of the cationic channels to cGMP \cite{RebrikKorenbrot04}.\\
Taken together, \calcium is now considered as the photoreceptor's internal messenger
for adaptation.  Supporting evidence comes from the fact that adaptation effects can
be provoked without light (c.f. p.130 in \cite{FainEtAlReview01}), by only
lowering the \calcium concentration, or that adaptation is suspended by clamping the
\calcium level to its value corresponding to darkness (p.126 in \cite{FainEtAlReview01}).\\
Beyond the level of the individual photoreceptor, further mechanisms related to
adaptation are effective, for example
network adaptation in interneurons and retinal ganglion cells (i.e. adaptation
is ``transferred'' beyond the receptive field of the actually stimulated cell,
e.g. \cite{VaqueroEtAl01,WeilterEtAl98,GreenEtAl75,BarlowLevick69}), and
discounting predictable spatio-temporal structures from the stimulus
by Hebbian mechanisms \cite{SnirBerWarBiaMei97,HosoyaEtAl05}.\\
%
%

%
\begin{table}
		\begin{tabular}{c||c|c}
		\hline	
		\emph{mechanism} 		&\emph{Ref.\cite{GrossHong03,GrossHong04}}	 & \emph{our approach}\\
		\hline	\hline
		light adaptation		& yes				& yes\\
		divisive gain control		& yes				& no\\
		\hline
	\end{tabular}
	\Caption[ModelOverview][Model overview][The table gives an overview over the mechanisms used in
	\new{the model of G\&H \cite{GrossHong03,GrossHong04} and our approach}.]
\end{table}
\def\Lumi[#1]{\mathcal{L}_\mathit{#1}}
\def\Vrest{V_\mathit{rest}}
\def\gLeak{g_\mathit{leak}}
\def\gExc{g_\mathit{exc}}
\begin{figure}[ht]
	\begin{center}
		\scalebox{0.55}{\includegraphics{\pics/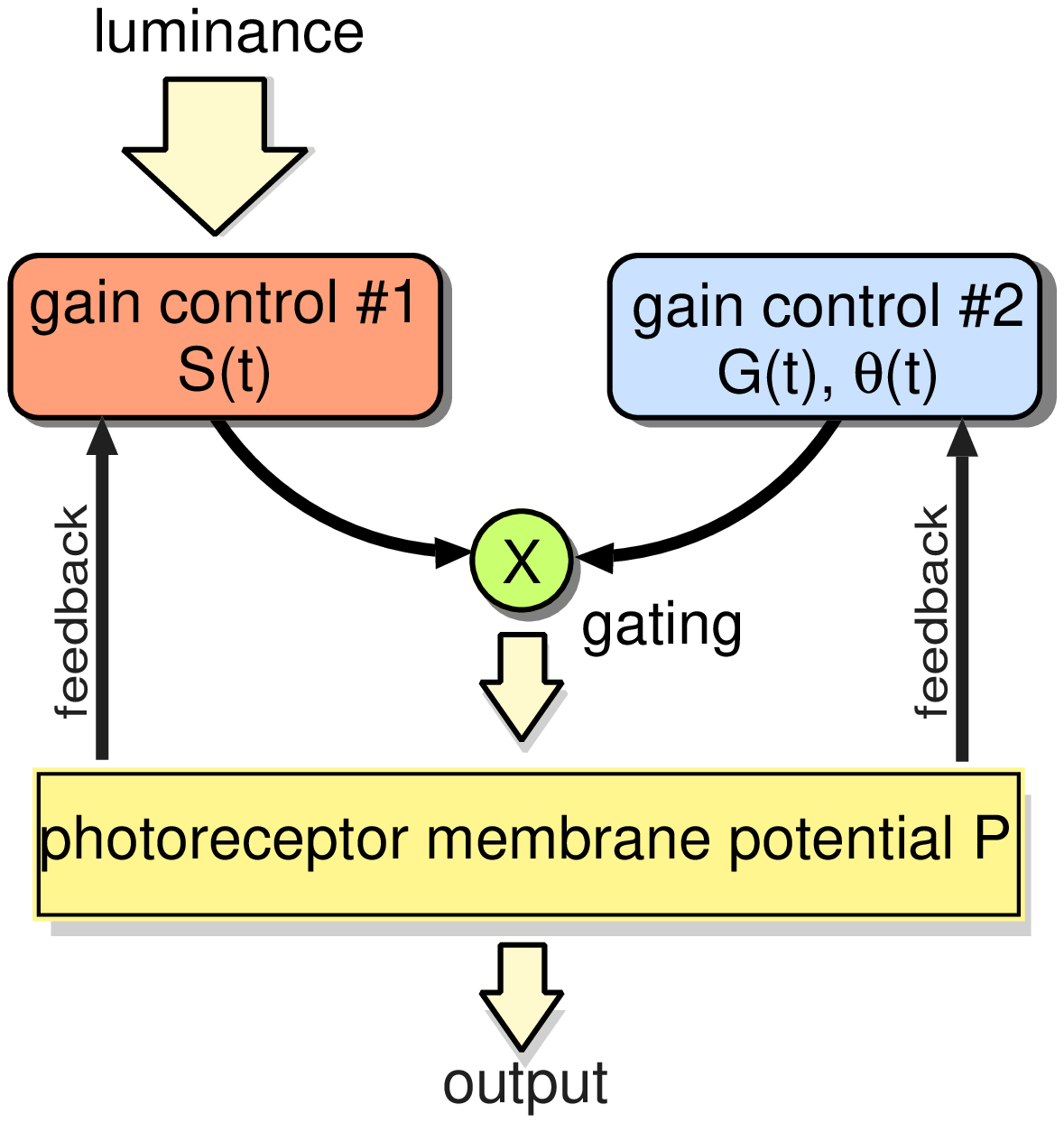}}
	\end{center}
	\Caption[ModelSketch][Model sketch][{A luminance distribution is subjected to
	a divisive ``gain control'' stage \#1 ($S(t)$, \eq[LumiCompression]).  At this
	stage, inhibition of luminance $\Lumi[]$ takes place as a function of increasing
	photoreceptor potential $P$.  The second gain control stage $G(t)$ can either amplify
	the signal $S(t)$ or attenuate it (equations \ref{GainControl},
	\ref{GainControlSwitch}, and \ref{threshold}).  Amplification of $S(t)$ occurs if
	the membrane potential $P$ falls below a threshold $\Theta$, and attenuation
	for $P>\Theta$ (\eq[GainControlSwitch]).
	Both ``gain control'' stages interact multiplicatively (denoted by the symbol
	``$\otimes$'', \eq[gating]) before providing excitatory input into the
	photoreceptor's membrane potential (symbol ``P'', \eq[photoreceptor]).
	The photoreceptor potential in turn feeds back into both of the ``gain control''
	stages.  At the same time, the photoreceptor potential represents the output
	of our model.}]
\end{figure}
\begin{figure}[ht]
	\begin{center}
		\scalebox{0.5}{\includegraphics{\pics/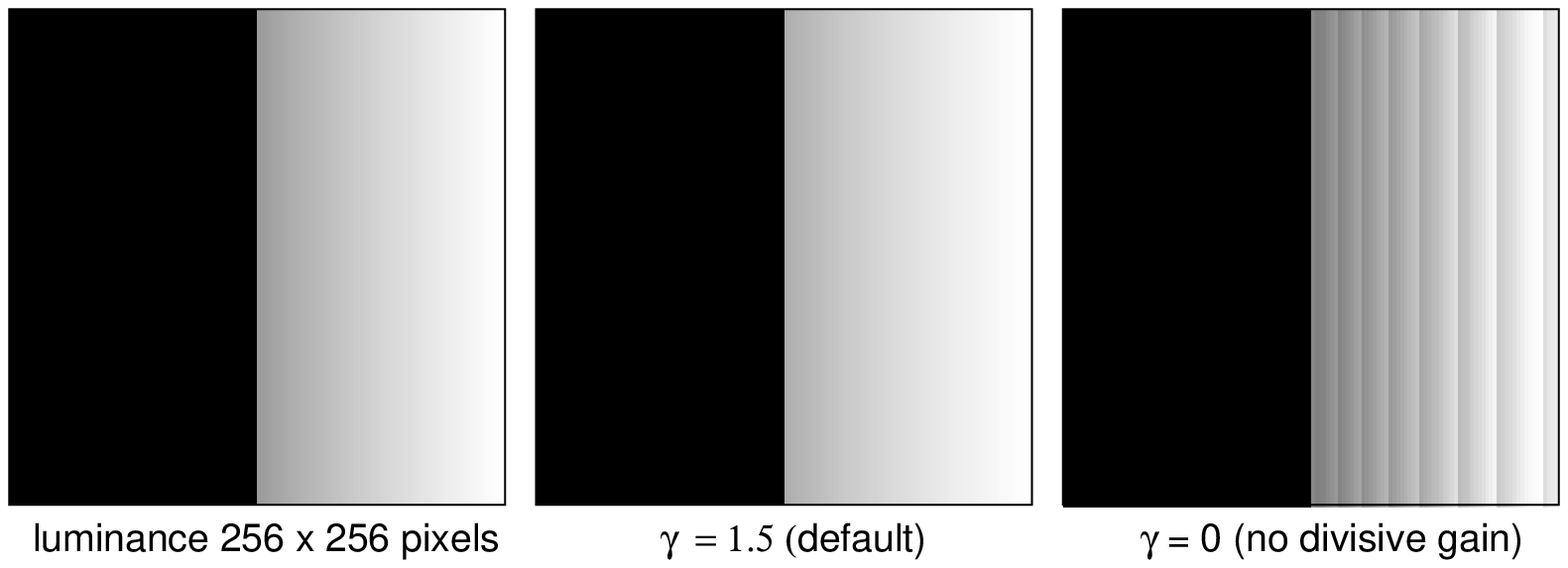}}
	\end{center}
	\Caption[gammaZero][Artifacts with a luminance ramp][{\emph{Left image}: The input
	$\Lumi[ij]$ , a luminance step with a superimposed luminance ramp (increasing
	linearly from left to the right).  \emph{Middle image}: With the default value
	$\gamma=1.5$ in \eq[LumiCompression] the adaptated image is correctly rendered
	and hardly distinguishable from the input.
	\emph{Right image}: Setting $\gamma=0$ causes the appearance of ripple
	artifacts in the adaptated image.  All results are shown at $t=250$ iterations.}]
\end{figure}
\begin{figure}[ht]
	\begin{center}
		\scalebox{0.275}{\includegraphics{\pics/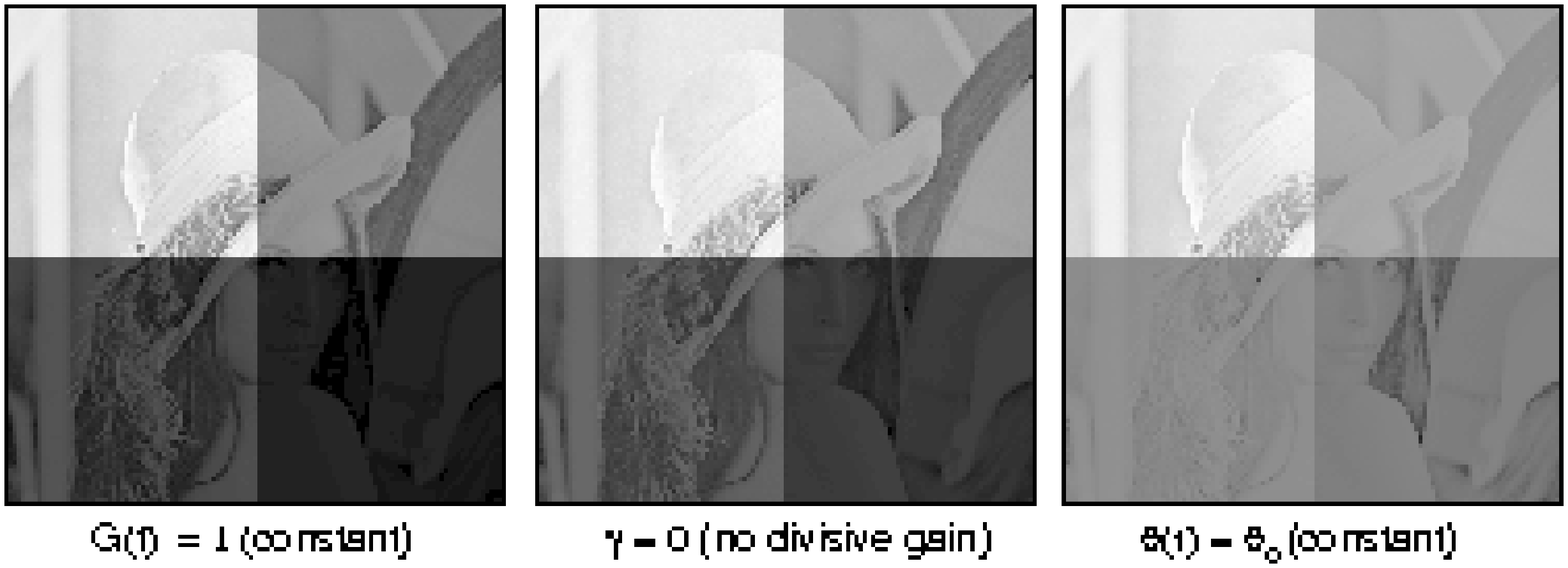}}
	\end{center}
	\Caption[Pathologies][Artifacts][{The results shown in this figure should be compared
	with \fig[LENA1]. \emph{Left image}: Setting the amplification constant to $G(t)=1$ in
	\eq[gating] diminishes adaptation (i.e., low luminance values are not pushed that high).
	Notice that in this case dynamical switching is made inoperative.
	\emph{Middle image}: Setting $\gamma=0$ in \eq[LumiCompression] has no effect on the
	\emph{natural} images we have tested, but causes strong ripple artifacts with luminance
	ramps as demonstrated in \fig[gammaZero]. \emph{Right image}:  Using a constant
	threshold $\Theta(t)=\Theta_0=0.25$ in \eq[GainControlSwitch] leads to strong
	saturation (or over-adaptation).  All results are shown at $t=250$ iterations.}]
\end{figure}
%
%
\section{Formal Definition of the Adaptation Dynamics} 
%
\Tab[ModelOverview] gives a brief comparison of \new{components,
and a sketch of our model is shown in \fig[ModelSketch].  In what
follows}, we give the formal introduction to our mechanisms
which are thought to provide an abstract view for adaptation as it
takes place in the outer segment of individual photoreceptors.\\
Let $\Lumi[ij]$ be a two-dimensional luminance distribution which
provides the input into our model.  For the purpose of the present
paper we assume that the model converges before changes in luminance
occur, that is $\partial\Lumi[ij](t)/\partial $t$=0$, where spatial
coordinates are denoted by $(i,j)$.  We assume that the input is
normalized according to $\epsilon < \Lumi[] \leq 1$, with $\epsilon$
is chosen such that $0 < \epsilon < \min_{i,j}\{\Lumi[ij]\}$.  Let $P$
denote the membrane potential of the photoreceptor, which is assumed
to obey the equation (the symbols $\gLeak$, $\gExc(t)$, and $V_{exc}$
are defined below)
\begin{eqnarray}\label{photoreceptor}
	\ddt[{P}]	= - \gLeak P(t) + \gExc(t)
					\left[ V_{exc}-P(t)\right]
\end{eqnarray}
An instance of the last equation holds for each position $(i,j)$, hence
$P\equiv P_\mathit{ij}(t)$ (in what follows indices were dropped for
brevity).  The excitatory saturation point (or reversal potential)
is defined by $V_{exc}$, and the leakage (or passive) conductance
is defined by $\gLeak$  (note that $V_{exc}$ represents an asymptote for $P$).
Both of the last constants are equal for all photoreceptor cells.
The default simulation parameters, as well as further simulation details,
can be found in table \ref{SimuPars}.  Notice that photoreceptors in fact
hyperpolarize in response to light (c.f. section \ref{adaptation}), whereas the last
equation makes a contrary assumption.  This assumption, however, implies
no loss of generality, since the model can equivalently be re-formulated
such that it hyperpolarizes with increasing intensity levels.\\
Excitatory input to the photoreceptor potential is given by the conductance
$\gExc\equiv g_\mathit{exc,ij}(t)$, which is defined by
\def\Gain[#1]{G_\mathit{#1}}
\begin{eqnarray}\label{gating}
	\gExc(t)	= \Gain[](t) \cdot S(t) 
\end{eqnarray}
where the process $\Gain[]\equiv \Gain[ij](t)$ interacts multiplicatively
with the light-induced signal $S\equiv S_\mathit{ij}(t)$ (such interaction
was previously referred to as mass action or gating mechanism, see
\cite{CarpenterGrossberg}).  For the signal $S$, we assume that its
efficiency for driving the photoreceptor's potential diminishes with
increasing potential~$P$:
\begin{eqnarray}\label{LumiCompression}
	S(t)	= \frac{\Lumi[]}{1+\gamma \cdot P(t)} 
\end{eqnarray}
The last equation in fact establishes a feedback mechanism which allows
the photoreceptor to regulate the strength of its own
excitatory input.  In addition, the excitatory drive of the photoreceptor
is also a decreasing function of increasing potential $P(t)$ by virtue of
the term ``$(V_{exc}-P)$'' (the driving potential) in \eq[photoreceptor]. 
Notice that if $\gExc$ was constant and sufficiently high, the driving
potential would make $P(t)$ saturate at  $V_{exc}$ (i.e., $V_{exc}$ is
asymptotically approached).  Therefore, both the excitatory input $\gExc$,
and the driving potential, decrease as $P(t)$ grows.  The motivation for
including \eq[LumiCompression] in our model was to eliminate ripple
artifacts seen with luminance ramps (\fig[gammaZero]).
With ``normal'' natural images, those artifacts did not appear to be a
major nuisance (\fig[Pathologies]; see also section \ref{DescriptionOfDynamics}).\\
The process $\Gain[](t)$ implements an amplification mechanism as follows: 
\begin{eqnarray}\label{GainControl}
	\tau_k \ddt[{\Gain[]}]	= -\Gain[] + \delta(t-t_0)
\end{eqnarray}
where $\delta(t-t_0)$ is the Dirac function which serves to impose the
initial condition $\Gain[](t=t_0)= 1$, given that $\Gain[](t)=0\ \forall\ t<t_0$
(notice that $\Gain[]\equiv\Gain[ij](t)$ as usual).
Simulations are assumed to start at $t_0=0$.  By virtue of the
index $k\in\{1,2\}$ associated with the time
constant $\tau_k$, the last equation describes \textit{two} distinct processes.
These processes are characterized by $\tau_1>0$ (making $\Gain[]$ decay with time),
and $\tau_2<0$ (leading to an increase of $\Gain[]$ with time).  The last equation
thus implements what we dubbed a ``dynamically switching gain control''.  But who
or what is switching $\Gain[]$ on (i.e. making it increase with $|\tau_2|$) or off
(i.e., making it decrease with $\tau_1$)?  The one or the other process is
invoked depending on whether $P$ exceeds a threshold $\Theta$ or not:
\begin{eqnarray}\label{GainControlSwitch}
	k=1 & \mathrm{if} & P(t)>\Theta(t)\\\nonumber
	k=2 & & \mathrm{otherwise}
\end{eqnarray}
This means that if the outer segment potential $P$ is below the threshold $\Theta$,
its input $\gExc(t)$ is amplified via \eq[LumiCompression].  The amplification
mechanism acts to diminish the integration time of luminance signals until
reaching the threshold $\Theta$, especially low-intensity signals.  Once the
threshold is exceeded, amplification is switched off (\fig[GainG_3d]).  In fact,
$\Gain[]$ decays rapidly then in order to avoid driving the outer segment potential
into saturation (which nevertheless may occur at sufficiently high intensity
values).  \new{With ineffective dynamical switching $G\equiv \mathrm{const.}$ adaptation
is severely deteriorated (\fig[Pathologies], first image)}.  Mathematically, the dynamic
switching mechanism avoids an unbounded growth of $G$.\\
Amplification proceeds until $P$ crosses a threshold. The threshold, however,
is not fixed, but is rather represented by a slowly decaying process on its own:
\begin{eqnarray}\label{threshold}
	\tau_\Theta \ddt[{\Theta}]	= -\Theta(t) + \Theta_0\cdot\delta(t-t_0)
\end{eqnarray}
The Dirac function $\delta(t-t_0)$ establishes the initial condition
$\Theta(t=t_0)= \Theta_0$, if $\Theta(t)=0\ \forall\ t<t_0$ (notice that
$\Theta\equiv\Theta_\mathit{ij}(t)$).  We like to emphasize that the
threshold $\Theta$ is not supposed to represent a firing threshold for
the photoreceptor.  It rather serves to implement
the dynamic switching behavior for turning the signal amplification on or
off.  The motivation for including a dynamical threshold in our model was the
elimination of artifactual contrasts inversion effects, and will be
explained in more detail in section \ref{DescriptionOfDynamics}.
\new{Furthermore, if a constant threshold were chosen, over-adaptation
would occur (\fig[Pathologies], last image)}.\\
Our simulations were evaluated at the moment when
$P_\mathit{ij}>\Theta_\mathit{ij}\ \forall\ (i,j)$.  This
is, however, not a steady-state, because the outer segment potential
continues to decay with $\gLeak$.  The results which are presented in
the figures \ref{MIT0} to \ref{PEPPERS5} therefore show snapshots of
the outer segment potential at exactly the moment when the last potential
value $P_\mathit{ij}(t)$ exceeded the threshold $\Theta(t)$ (i.e., $(i,j)$
corresponds to the position with the lowest intensity value in the input).
One may ask why we gave preference to a dynamical formulation of our
model over steady-state equations.  Intuitively, steady-state solutions
cannot capture the full behavior revealed by the model.  For example,
the steady-state solution (as defined by $d\Theta/dt=0$) of the last
equation is zero, and, depending on $k$, the steady-state solution of
\eq[GainControl] is infinity ($k=2$) or zero ($k=1$).
%
\begin{table*}[t]
	\begin{minipage}{.55\textwidth}\small
		\begin{tabular}{c||c|c|l}
			\hline	
			\emph{parameter} 	&\emph{value}	& \emph{equation} 		& \emph{description}\\
			\hline	\hline
			$\gLeak$		&0.05		& \ref{photoreceptor}		& leakage conductance\\
			$V_{exc}$		&1		& \ref{photoreceptor}		& synaptic battery\\
			$\gamma $		&1.5		& \ref{LumiCompression}		& divisive gain\\
			$\tau_1$		&0.7213		& \ref{GainControlSwitch}	& damping time constant\\
			$\tau_2$		&-40.4979	& \ref{GainControlSwitch}	& amplification time constant\\
			$\Theta_0$		&0.25		& \ref{threshold}		& initial threshold value\\
			$\tau_\Theta$		&39.4949	& \ref{threshold}		& threshold decay time constant\\
			\hline
		\end{tabular}
	\end{minipage}
	\begin{minipage}{.40\textwidth}
		\begin{center}
		\Caption[SimuPars][Simulation details][{The table is self-explanatory.  For the integration of
		\eq[photoreceptor] a fourth-order Runge-Kutta scheme was used with an integration time step of
		0.01.  The remaining differential equations were integrated with Euler's method with an integration
		time step of one.  Notice that the integration step sizes were not adjusted to match physiological
		time scales.}]
		\end{center}
	\end{minipage}
\end{table*}
%
%
\begin{figure}[ht]
	\begin{center}
		\scalebox{0.75}{\includegraphics{\pics/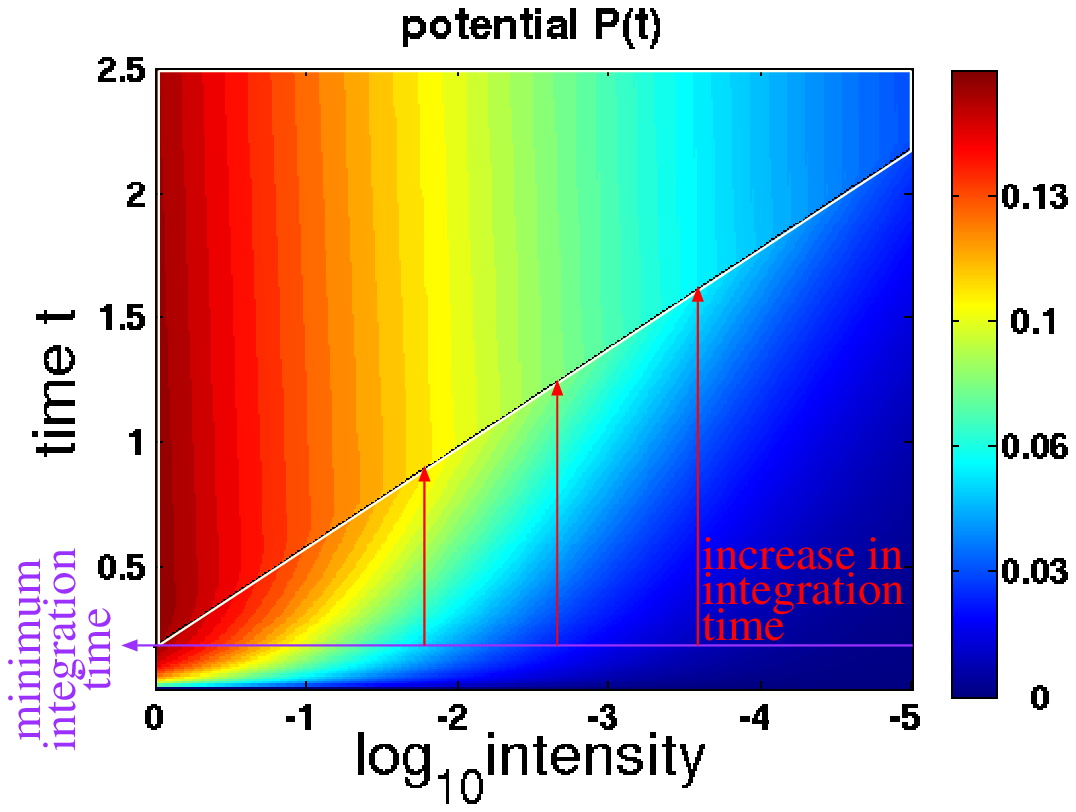}}
	\end{center}
	\Caption[PhotoP_3d][Photoreceptor potential][{The photoreceptor
	potential $P$ (\eq[photoreceptor]) is plotted as a function of time ($t=0$ to
	$250$ iterations) and input intensity ($\Lumi[]\in\{10^{0},10^{-1},...\,,10^{-5}\}$.
	The photoreceptor amplitude is color-coded (colorbar). 
	``Convergence'' occurs when the photoreceptor potential $P$ exceeds
	a threshold $\Theta$, and corresponds to the area over the diagonal line. 
	The minimum integration time is delinated by the horizontal line at the
	bottom.  With decreasing luminance, one observes an increase in integration time until
	``convergence'' is reached (as illustrated by the red arrows pointing to the plateau).
	A similar increase in integration time with decreasing stimulus intensity 
	levels is also known from the retina, and is expressed as Bloch's law of
	temporal integration.  Bloch's law relates the threshold for seeing a stimulus
	to stimulus duration (i.e. integration time) and stimulus intensity: the product
	of stimulus duration and stimulus intensity equals a constant within a so-called
	critical time window.  Bloch's law is especially prominent for scotopic vision.}]
\end{figure}
\begin{figure}[ht]
	\begin{center}
		\scalebox{0.45}{\includegraphics{\pics/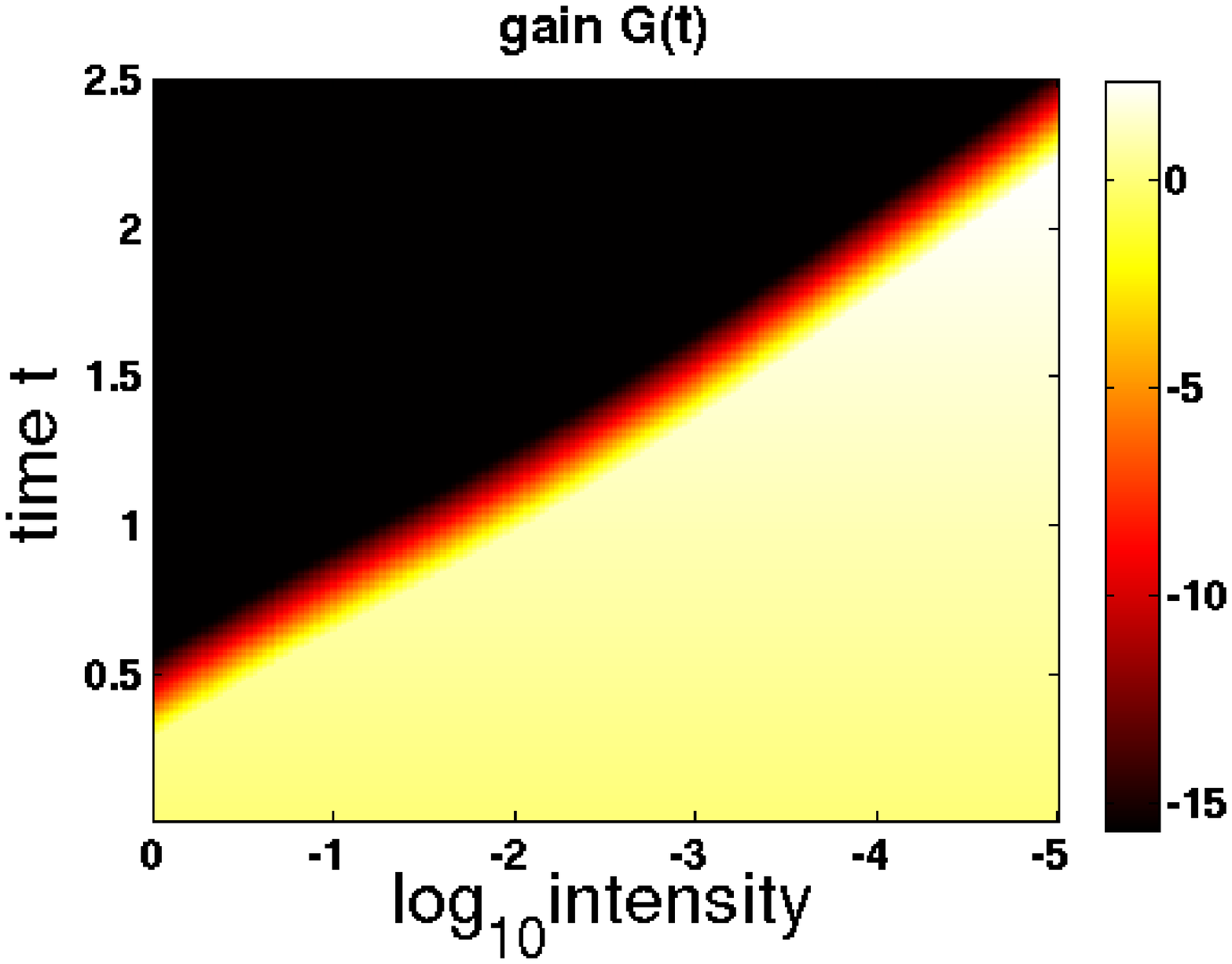}}
	\end{center}
	\Caption[GainG_3d][Dynamics of the ``switching'' gain control][{The same as
	in \fig[PhotoP_3d], but here the dynamics of the signal amplification variable
	$G(t,\Lumi[])$ (\eq[GainControl]) is visualized.  The bright (dark) area on the
	bottom (top) indicates where the gain control is switched on (off).  Notice
	that the switching occurs rather fast around the red area.  The switching
	area resembles a blurred line - compare it to the diagonal line delineating
	the convergence plateau in \fig[PhotoP_3d].}]
\end{figure}
\begin{figure}[ht]
	\begin{center}
		\scalebox{0.325}{\includegraphics{\pics/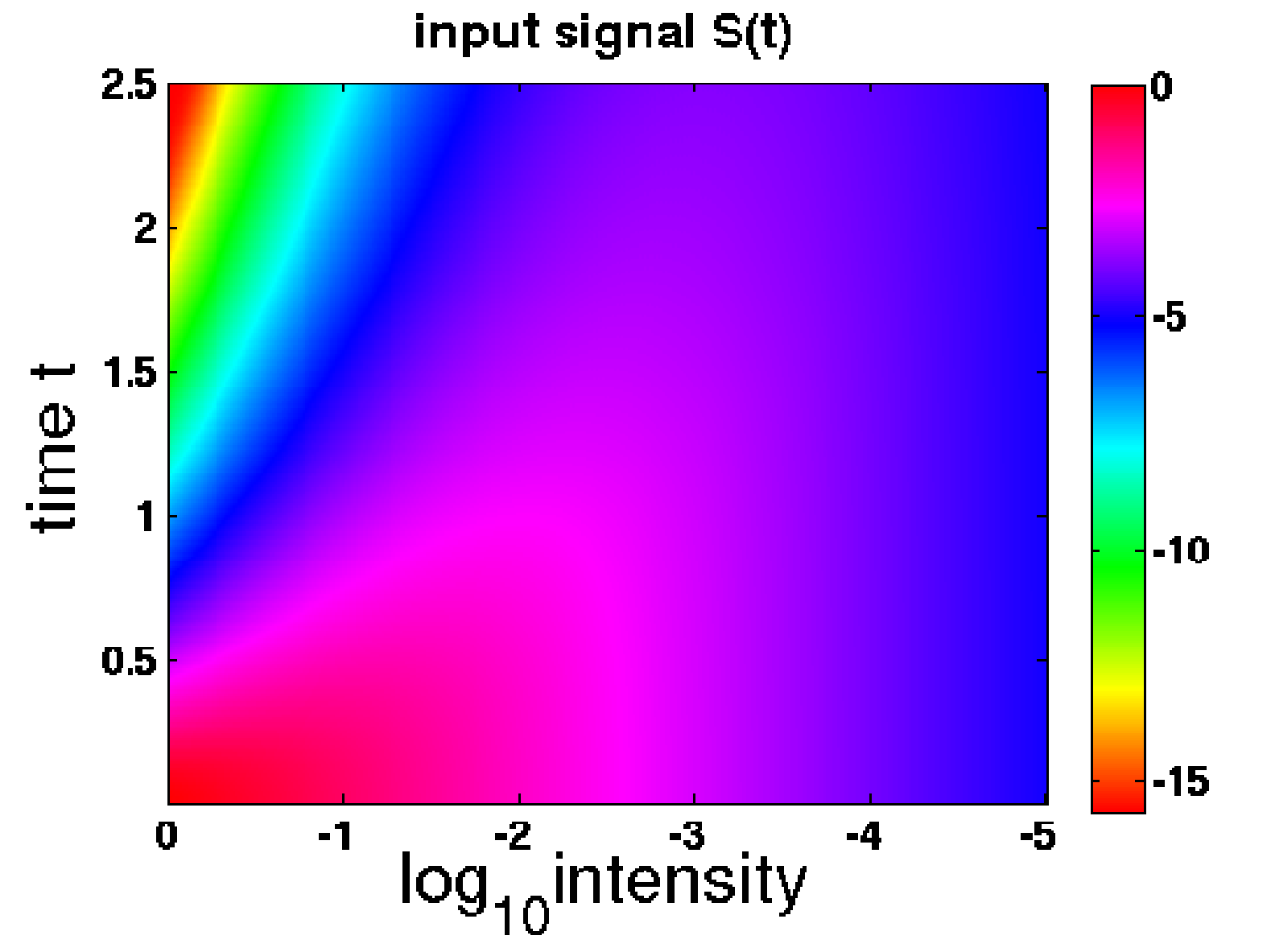}}
	\end{center}
	\Caption[GainI_3d][Input signal][{The same as in \fig[PhotoP_3d],
	but here the dynamics of the input signal $S(t,\Lumi[])$ is
	visualized (\eq[LumiCompression]).}]
\end{figure}
%
\section{\label{DescriptionOfDynamics}Description of the Adaptation Dynamics} 
%
What does the adaptation dynamics defined by equations \ref{photoreceptor} to \ref{threshold}
look like? The process obviously integrates the activity generated by an input image $\Lumi[]$,
via the photoreceptor membrane potential $P$.  The integration proceeds until $P$ exceeds the
threshold $\Theta$.  At this point the integration process decelerates exponentially with a
time constant $\tau_1>0$, since the corresponding solution to \eq[GainControl] is
$\Theta(t)=\exp(-t/\tau_1)$.  The dynamics of $P$ is shown in \fig[PhotoP_3d]: luminance
values that vary over 5 orders of magnitude are mapped onto roughly two orders of output
magnitude in a way that contrast relationships of the input are preserved.  Moreover,
the process converges rather fast.  Even for the smallest input intensities, convergence
is reached at about 200 iterations.  This fastness is a consequence of the dynamic
switching process, which increases signal amplification $\Gain[]$ until $P$ exceeds $\Theta$
(doing so reduces the integration time especially for weak luminance signals).
Since this process (\eq[GainControl]) \emph{per se} would grow in an unbounded fashion,
one may question its physiological plausibility.  But as long as $\epsilon>0$, or dynamically
varying noise is present in the model, eventually all luminance values reach threshold in
finite time, and as a consequence $G$ (\eq[GainControl]) switches from amplification to
attenuation.  This is to say that for $k=2$ the process $G$ is bounded mathematically
from above.
Furthermore, numerical experiments demonstrate that $G$ does not adopt excessively
high values (see \fig[GainG_3d]) \footnote{If $P(t)\leq\Theta(t)$, the sub-threshold
gain obeys $G(t)=\exp(t/|\tau_2|)$.  Assuming $t=250$ iterations and using 
$|\tau_2|=40.5$ (see \tab[SimuPars]) we get $G(t=250) \approx 479.55$ as maximum
amplification.}.\\   
Nevertheless, a suitably parameterized and asymptotically bounded process for
substituting $G$, rather than a sharply cut exponential (as it is implemented by
the equations \ref{GainControl}, \ref{GainControlSwitch} and \ref{threshold}), would
perhaps better reflect physiological reality -- but for the moment we set aside
plausible functions to keep the model concise.\\
Why should the threshold $\Theta$ drop with time? Imagine that we fix $\Theta$ to
some constant value.  In that case, all luminance values are integrated until they
all reach the \emph{same} threshold.  This means that the integration process would
establish a common level for bright and dark luminance values, what in the best of
all cases would lead to a strong reduction of contrasts with respect to the input
(\fig[Pathologies], last image).  But there is yet another, more technical point,
to this.\\
Consider a pair of luminance values, one brighter than the other.
Since the integration process proceeds with fixed time steps (and exponentially increasing gain),
we may choose both luminance values such that they exceed the fixed threshold in a way that the
previously dark luminance value leaves \emph{more} super-threshold activity than the bright value
(the brighter value must have exceeded threshold at some former time step, and thus its activity $P$
already has decayed somewhat due to the passive leakage conductance $\gLeak$ in \eq[photoreceptor]).
In other words, when decoding the photoreceptor potential $P$, the dark value would
suddenly appear brighter than the original bright value.  Such ``contrast inversion'' artifacts
are avoided  with a  threshold that decreases with time.  Thus, the dynamic threshold process
(\eq[threshold]) acts to preserve contrast polarities (notice that the threshold process
asymptotically approaches zero).\\
Yet another type of artifact may emerge as a consequence of the exponentially increasing
amplification signal $G$, most likely due to amplification of numerical noise while
integrating the differential equations.
With certain luminance distributions, especially with luminance ramps, step-like or
ripple-like structures may appear when $P$ is read out (of course the ripples are absent from the
input, c.f. \fig[gammaZero]).
Those artifacts are counteracted by the additional gain control mechanism (\eq[LumiCompression]).
Its net effect is to continuously decrease the integration step size for \eq[photoreceptor] as the potential
$P$ grows.  This effect gets especially prominent for high luminance values (see \fig[GainI_3d]).\\
\begin{figure}[ht]
	\begin{center}
		\scalebox{0.275}{\includegraphics{\pics/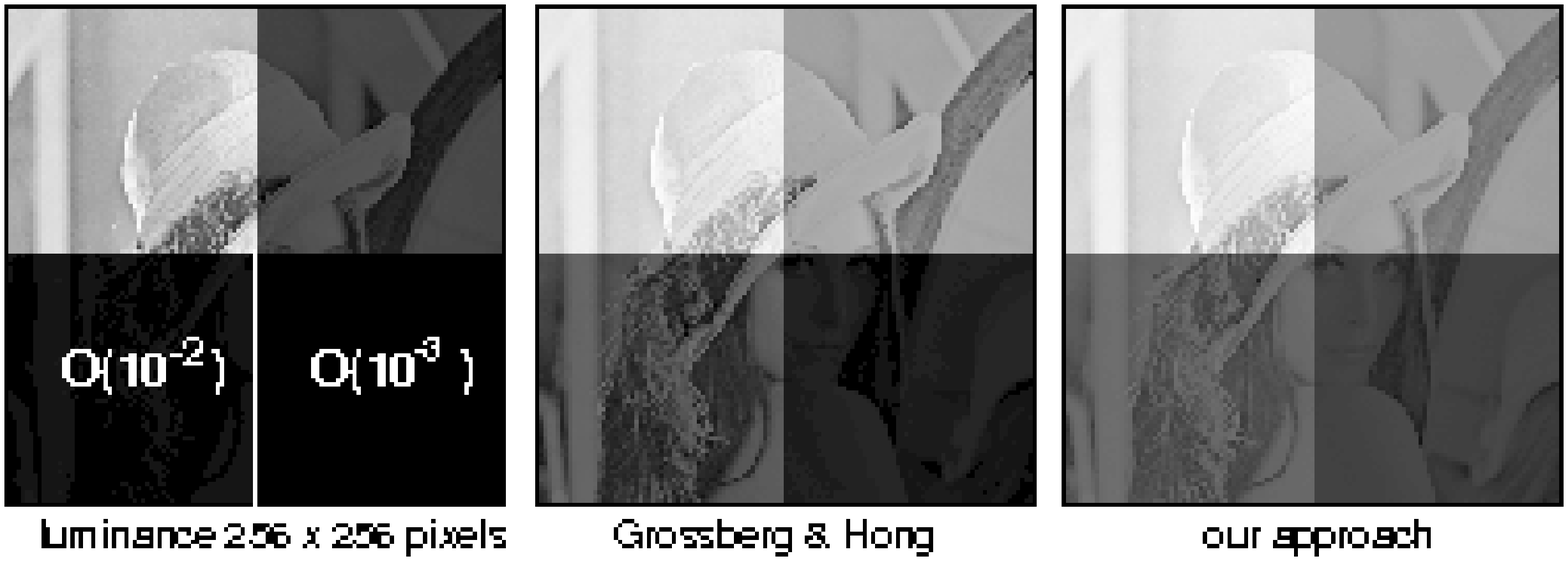}}
	\end{center}
	\Caption[LENA1][Tiled Lena image][{The original
	Lena image (with luminance values between 0 and 1, see \fig[LenaPeppersLumi])
	was subdivided into four tiles, and tiles were multiplied with $10^{0}$,
	$10^{-1}$, $10^{-2}$, and $10^{-3}$, respectively.  In the input (leftmost image),
	both of the lower tiles are displayed in black.  The order of magnitude of the
	corresponding luminance range is indicated with the black tiles.}]
\end{figure}
\begin{figure}[ht]
	\begin{center}
		\scalebox{0.275}{\includegraphics{\pics/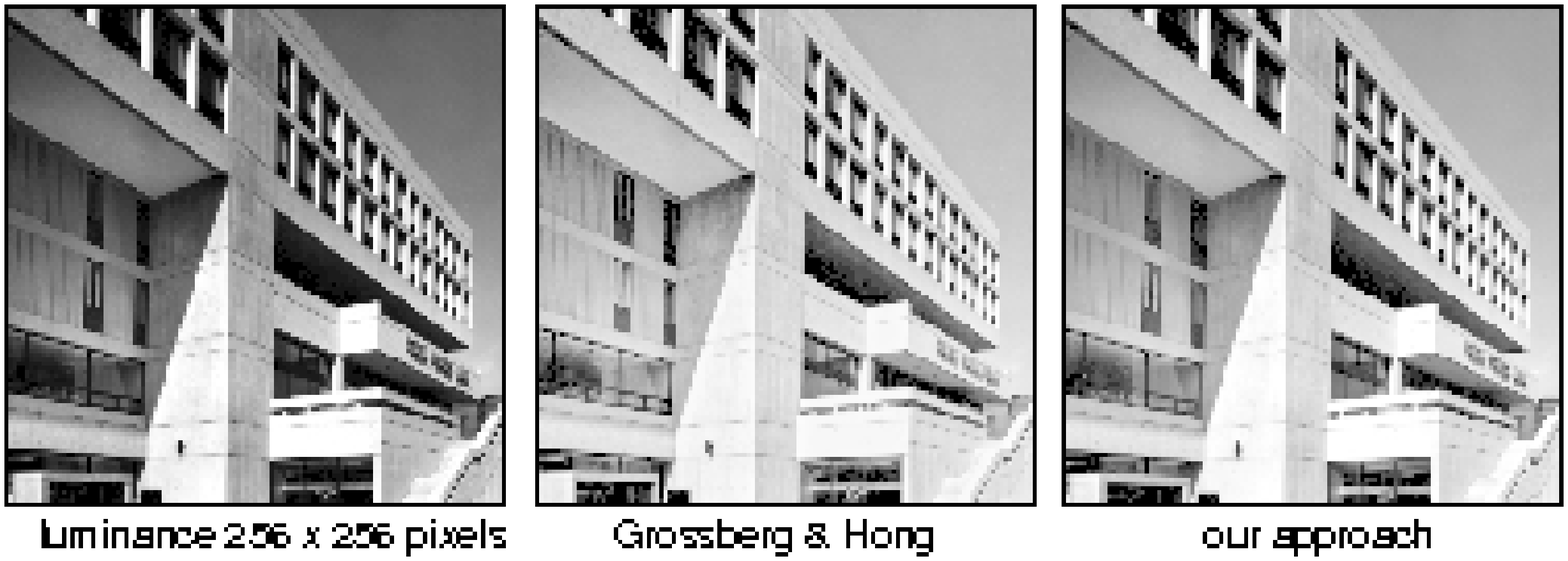}}
	\end{center}
	\Caption[MIT0][MIT image][{The first image shows the input
	image, with luminance values originally varying from 0 to 255.
	The input image was normalized such that the maximum intensity
	value was 1, and the minimum 0.  Subsequently, all zero
	luminance values were substituted by $\epsilon=(1/255)/2$.
	The second
	image shows the result obtained with the method described in
	\cite{GrossHong03,GrossHong04} (500 iterations).  The last
	image was obtained with our approach (150 iterations;
	convergence occurred within simulation time).  Both
	results show the cone's membrane potential.}]
\end{figure}
%
\section{\label{Results}Results of Numerical Experiments} 
%
What should one expect from a ``good'' adaptation mechanism? It should map
luminance values, which can be distributed over several orders of magnitude,
onto a fixed target range of, say, one or two orders of magnitude.  In this
way, images with a high dynamic range could be visualized with a normal
computer monitor.  If we tried a direct visualization of a high dynamic
range image without applying any adaptation,
we could just see the luminance patterns of the first one or two orders of
magnitude, while all smaller luminance values would be displayed in black
(see \fig[LENA1]; notice that the optic nerve has a similar transmission bandwidth).
Additionally, a ``good'' adaptation mechanism should leave unchanged an input
image which does only vary over one or two orders of magnitude.  Or at
least leave such an image as far unchanged as possible.  Contrast strength
should ideally be preserved.  Put another way, compression effects that
are introduced by the adaptation mechanism should be minimized.\\
We compare the results of our mechanism with one proposed by
\cite{GrossHong03,GrossHong04} (subsequently denoted
by ``G\&H'') \footnote{We implemented equations
A3 to A8 from ref.\cite{GrossHong04}, and integrated their model
over 500 iterations with Euler´s method, where a integration step
size of 0.01 was used.}.\\
In order to assure that, at some time, $P(t)_{ij}>\Theta(t)$ at all
positions $(i,j)$, zero values of the original luminance distribution
were substituted by the half of the second smallest luminance value, that is
$\epsilon=0.5*min\{\Lumi[ij] : \Lumi[ij]>0\}$, if not otherwise stated.
We used standard benchmark images of size $256 \times 256$ pixels as inputs
$\Lumi[]$.\\
\Fig[MIT0] show the results with the MIT image, where the result obtained
with our method is slightly less saturated than the one obtained with G\&H's
method.\\
In order to better explore the performance of the two methods, we superimposed
the original test images with artificially generated illumination patterns.
In \fig[MIT2], the MIT-image was multiplied with a luminance ramp to simulate
an illumination gradient.  In the latter case, the result from G\&H is less
saturated than ours.\\
\begin{figure}[ht]
	\begin{center}
		\scalebox{0.275}{\includegraphics{\pics/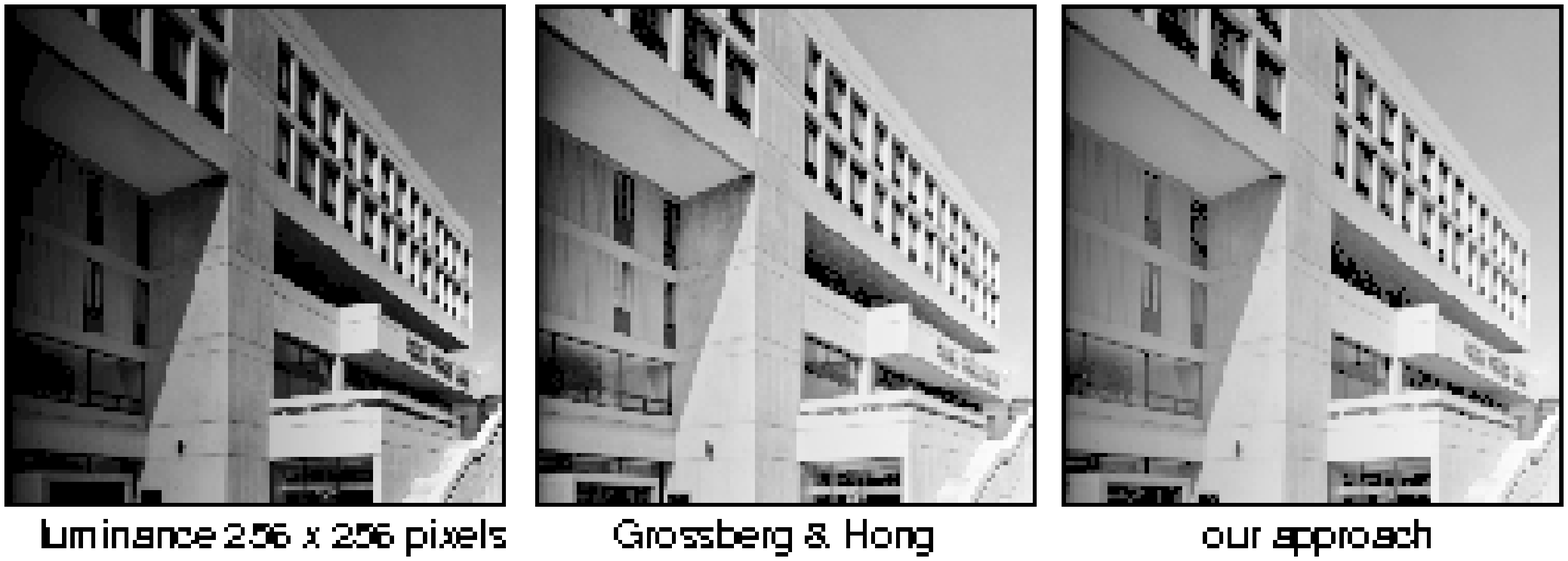}}
	\end{center}
	\Caption[MIT2][MIT image with overlying luminance ramp][{The original
	MIT image (see \fig[MIT0]) was multiplied with a luminance ramp which
	linearly increases from left (intensity 0) to the right (intensity 1).}]
\end{figure}
In \fig[LENA1], the original image (shown in \fig[LenaPeppersLumi]) was
subdivided in four ``tiles'', where within each tile luminance values
vary over a different order of magnitude.  This test image
mimics a situation where the range of luminance values within a
scene varies over four orders of magnitude.  Both methods push
luminance values sufficiently high such that details in the darkest
tile are rendered visible (where our method yields an overall more
brighter result -- and hence the darkest patch is better visible).
Thus, four orders of magnitude of input range are mapped onto two
orders of magnitude available for visualization, a situation that
is similar to situations which are met by the retina.\\
In the last example, we created an artificial high-dynamic range image (\fig[PEPPERS5])
from the original ``Peppers'' image (\fig[LenaPeppersLumi]).  In this case, our method
produces a slightly brighter result compared with G\&H: the result generated
with G\&H's method has harder contrasts.\\
\begin{figure}[ht]
	\begin{center}
		\scalebox{0.275}{\includegraphics{\pics/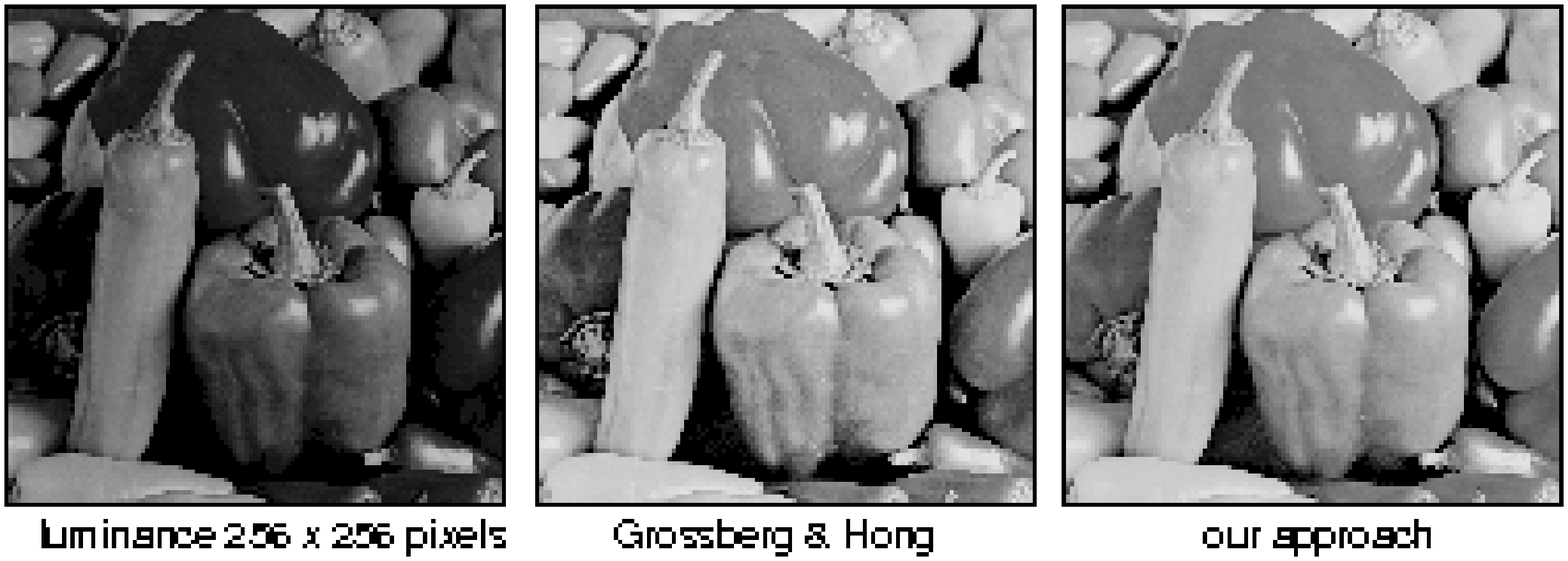}}
	\end{center}
	\Caption[PEPPERS5][Power-law-stretched Peppers image][{Luminance values of
	the original Peppers image (see \fig[LenaPeppersLumi]) were raised to the
	power of 4 to create a high dynamic range image.}]
\end{figure}
\begin{figure}[ht]
	\begin{center}
		\scalebox{0.275}{\includegraphics{\pics/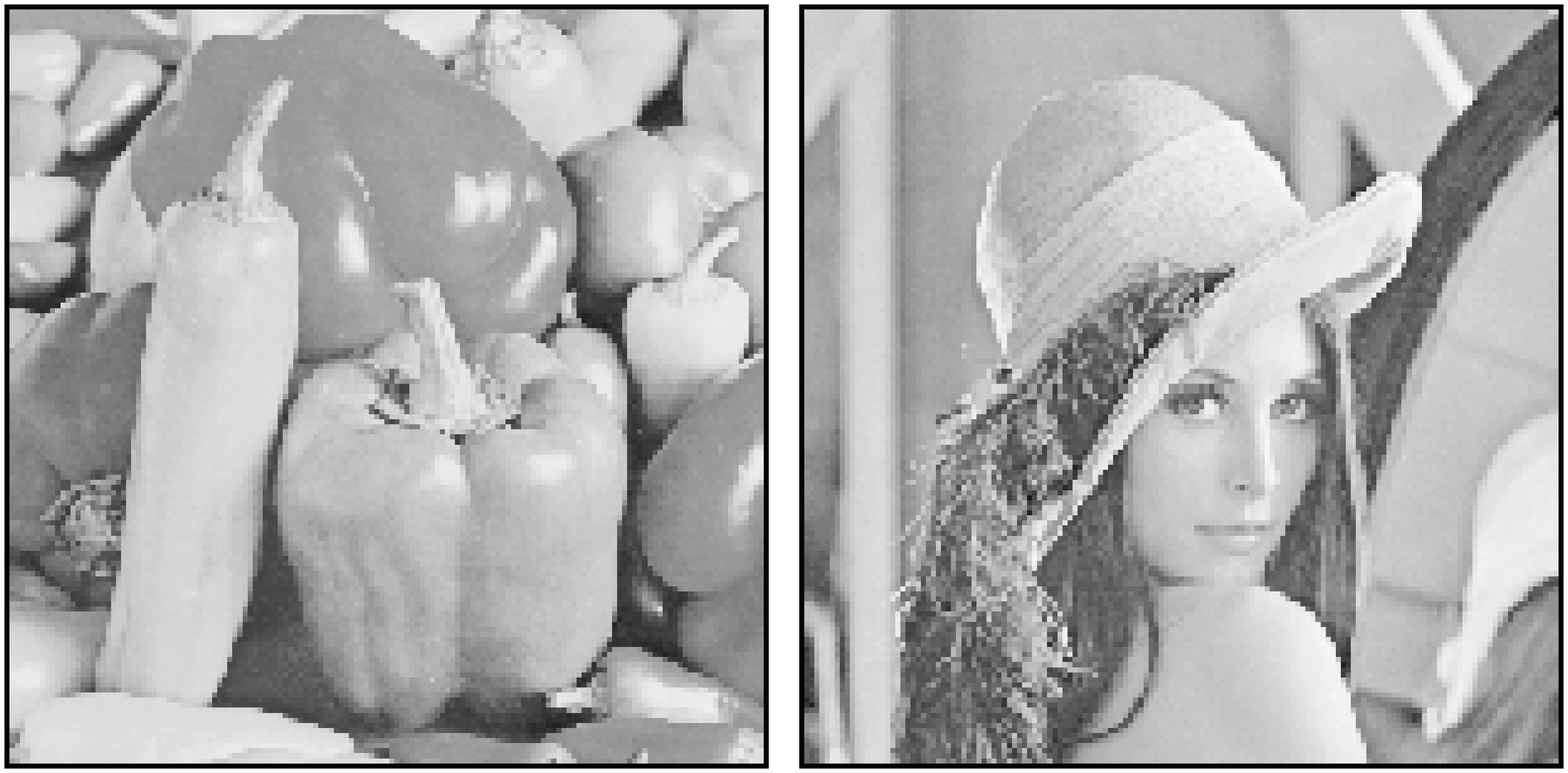}}
	\end{center}
	\Caption[LenaPeppersLumi][Original ``Lena'' and ``Peppers'' image][{These images
	are shown for comparing them with the results presented in figure \ref{LENA1}
	and \ref{PEPPERS5}, respectively.}]
\end{figure}
We conducted further simulations where we set $\Lumi[ij]\leftarrow P_\mathit{ij}$ after convergence, and
re-started the simulation.  The results did not change, indicating that the model's state after converging
the first time already corresponds to a steady-state solution.
%
%
\section{Model behavior with parameter changes} 
%
The parameters of our model can be tuned according to the expected numerical
range of luminance values.  In this way, compression effects in the output
are reduced, what can lead to the generation of visually more pleasing results.\\
Increasing the value of $\gamma$ (\tab[SimuPars]; \eq[LumiCompression]) reduces
the overall compression of the input at the cost of low-intensity regions.  This
is to say that low-intensity regions will appear darker, and regions with
higher intensities will be rendered with somewhat improved contrasts.  A
similar effect results, albeit more intense, when increasing the
threshold decay time constant $\tau_\Theta$ (\eq[threshold]).
Decreasing the initial threshold value $\Theta_0$ (\eq[threshold]) will
slightly increase overall brightness and compression, respectively.
The model behavior is quite robust against changes in the damping
time constant $\tau_1$, since this mechanism is backed up by the
signal gain control stage (\eq[LumiCompression]).  Nevertheless, variations
in the value of the amplification time constant $\tau_2$ bears strongly on the
results: a decrease improves greatly the adaptation behavior, but if $\tau_2$
is set too low artifacts may occur, such as contrast polarities being reversed
with respect to the input.
On the other hand, if $\tau_2\rightarrow\infty$, no adaptation at all takes place.
In future versions of our approach this influential parameter could be
set automatically as a spatially varying function of the structures in the input image.\\
%
\section{\label{epsilon}``This Thing Called Epsilon''...} 
%
As it turned out, a ``smart'' choice of $\epsilon$ can even improve
the contrasts in the visualization of the results.  Because for displaying,
each image is normalized to occupy the full range of available gray levels,
if $\epsilon$ is too small with respect to the second smallest luminance
value, it gets not sufficiently pushed by the adaptation process, such
that in the adapted image the difference between the smallest and the
second smallest value is too big.  As a consequence, many of the darker gray
levels are not used (if we assume a linear mapping of activity to gray levels),
what leaves less gray levels for displaying the other (higher) luminance values.
Hence, the contrasts in the displayed image will be reduced.
Ideally, $\epsilon$ should depend in some way on how dark the input image
is perceived by a human observer.  Finding an adequate function that
automatically sets the value of $\epsilon$ would be an interesting topic
for future research.
%
\section{\label{Conclusions}Discussion and Conclusions} 
%
We presented a novel theory about the adaptational mechanisms in retinal
photoreceptors.  Our theory is abstract in the sense that we did not attempt
to identify model stages with components of the phototransduction cascade
(as outlined in section \ref{adaptation}).  Nevertheless, one is tempted to draw
corresponding parallels between our model and physiological data.
\new{In the transduction cascade} there are (at least) two sites of amplification:
the serial activation of transducins by the active form of rhodopsin Rh$^*$, and
the hydrolysis of cGMP by phosphodiesterase.
An amplification of the signal takes also place in our model by virtue of $G$
in \eq[GainControl].  Furthermore, \calcium  constitutes a messenger for adaptation.
In contrast, there is no corresponding variable for describing the concentration
of \calcium  in our model.  Nevertheless, the membrane potential $P$ subserves two
different purposes.  First, it corresponds to the output of the photoreceptor.
Second, it constitutes a feedback signal  that acts to control signal amplification
-- and hence the adaptation process.
As \calcium is known to be linearly related to the membrane potential, it seems
reasonable to consider $P$ as a lumped-together description for both the
membrane potential and the \calcium concentration.\\
Indeed, one can draw further
parallels.  In our model, signal amplification stops as soon as the membrane
potential exceeds a threshold, in order to counteract saturation effects
(\eq[GainControlSwitch]).  This process is reminiscent on the binding of
arrestin to phosphorylated Rh$^*$, leading to a complete inactivation of
the photopigment, and thus to a ceasing of the transduction cascade.\\
In our model, there is yet another way to counteract saturation effects, by means
of the divisive inhibition stage (\eq[LumiCompression]).  This process
can be compared to the interaction of \calcium with the visual cycle, which
causes an acceleration of
the rate of Rh$^*$ phosphorylation \cite{Kawamura93,ChenEtAl95,KlenchinEtAl95}.
This interaction is brought about by the \calcium-binding protein recoverin,
and decreases the lifetime of Rh$^*$.  As a consequence, less cGMP will be
hydrolyzed upon absorption of a photon \cite{CalvertEtAl02}.\\
On the technical side, computer simulations demonstrated that our approach
is on a par with a recently proposed model from Grossberg and Hong
\cite{GrossHong03,GrossHong04} (``G\&H''). However, several crucial
differences exist between their approach and ours.\\
First and above all, the critical stage for adaptation in G\&H's approach
consists of the feedback provided by electrically coupled horizontal cells.
Light adaptation through the outer segment can be decoupled from the actual
adaptation dynamics, and hence may be considered as a pre-processing step
in their model.\\
Remarkably, our approach achieves similar adaptation results \emph{without}
incorporating the horizontal-to-cone feedback loop.  This prediction is
consistent with physiological data, as cone photoreceptors can decrease
their sensitivity over about 8 log units of background intensity \cite{Burkhardt94}.
Moreover, feedback from horizontal cells may even further improve adaptation.
Since we have seen, on the other hand, that contrasts are reduced as a consequence
of the dynamic range compression, one may speculate that feedback from horizontal
may also compensate for this effect, by re-enhancing contrasts.  Notice that
contrast enhancement is tantamount to center-surround interactions.  Because
adjacent horizontal cells of the same type are fused by gap junctions, their
feedback will influence the membrane potential of neighboring cones within
some radius of the actually stimulated photoreceptor.  In this way, the
antagonistic receptive field structure is created in bipolar cells.
But then bipolar cells represent a contrast-enhanced signal of the photoreceptors.
Therefore, neurophysiological data are consistent with our ideas.\\
Both models are similar complex with respect to parameter spaces.
G\&H's approach has some 10 parameters, whereas ours has
7 (plus the $\epsilon$).  Although we did not carry out a detailed
analysis of computational complexity, the respective model structures
suggest that the G\&H model is computationally more demanding.  The
latter fact seemed to be confirmed with our simulations on a serial
computer, where our model converged in a fraction of the time that
was necessary to achieve comparable results with the G\&H model \footnote{In our
implementation of the G\&H model we used the steady-state equations where possible,
and also the long-range diffusion mechanism as proposed by the authors.}\\
\new{Similar to the G\&H model, another approach \cite{GrossBrajovic03}
is also motivated by the observation that strong contrasts usually
indicate reflectance changes in natural scenes, as opposed to intensity
variantions due to changes in illumination.
The approach from \cite{GrossBrajovic03}, however, has no stage for luminance
adaptation, and only computes an ``anisotropically-like'' smoothed version
of the image, which is used for exerting divisive gain control directly on
intensity values (c.f. \tab[ModelOverview]).  The lateral connectivity between cells that
form the diffusion layer is controlled by inverse Weber contrasts.  Hence, both
strong and weak contrasts in the original image may affect the degree of smoothing.
Simulation results obtained with our implementation of Gross' and Brajovic's
approach revealed strong boundary enhancement if tuned such that the adaptation
was comparable to the other two methods. This suggests that the signal transduction
characteristics of Gross' and Brajovic's approach is high-pass}.\\
Our model, perhaps with different parameter values, should as well be useful
for displaying high-dynamic range images, or synthetic aperture radar images.
This is a topic that will be pursued with future research.  Further interesting
questions address the incorporation of feedback from horizontal cells, and
possibly of reset mechanisms for the threshold process, in order to extend
our model's processing capacities to image sequences.
\begin{acknowledgments}
M.S.K. was supported by the \emph{Juan de la Cierva} program of the Spanish
government.  The authors acknowledge the help of two anonymous reviewers, whose
comments contributed to improve the first draft of this manuscript.  Further
support was provided by the MCyT grant TIC2003-00654.
 \end{acknowledgments}
%

%
\end{document}